# Nonlinear Cyclotron-Resonance Accelerations by a Generalized EM Wave

AKIMOTO, Kazuhiro and HOJO, Hitoshi [1)]

*School of Science and Engineering, Teikyo University, Utsunomiya, Japan*

1) *Plasma Research Center, University of Tsukuba, Tsukuba, Japan*

E-mail: akimoto@ees.teikyo-u.ac.jp

**Abstract**

Particle accelerations by a 1D, EM, dispersive pulse in an external magnetic field are investigated. It is found that the well-known cyclotron resonance may be classified into three regimes as the length and/or the amplitude of the pulse are varied. Namely, as the pulse amplitude increases, the transit-time cyclotron-resonance acceleration [CRA] evolves to phase trapping, and reflect particles. The amplitude and wave dispersion as well as the pulse length strongly affect those accelerations. The interesting phenomena of quantization of resonance velocities in between the two regimes are also investigated. This new mechanism may lead to wave amplification at some discrete frequencies other than the cyclotron frequency.

## 1. Introduction

Accelerations of charged particles by a sinusoidal EM wave have been studied extensively in the past.[1] Here, particle accelerations by a 1D, EM, dispersive wave packet in a magnetic field mainly through the cyclotron resonance are investigated. It is assumed that the pulse has a Gaussian profile with the following generalized linearly polarized electric field (and its self-consistent magnetic field) directed perpendicular to the z-axis along which it propagates:

$E_x(z,t) = E_0 \, e^{-\{(z-v_g t)/l\}^2 + i(k_0 z - \omega_0 t + \theta)}$ ; here, $E_0$, $z$, $v_g$, $t$, $l$ are the amplitude, the position, the central group-velocity, the time, and a measure of the pulse-length, respectively, and $\theta$ is the phase constant; furthermore, $k_0$ ($\omega_0$) is the wave number (angular frequency) of the carrier wave. This kind of EM pulse may approximate a part of a parallel shock wave located upstream from the earth's bow shock and/or the recently found EM solitons formed during laser fusion experiments. In the linear regime of the cyclotron resonance acceleration the velocity gain is relatively small, and the beam-like particles with only longitudinal initial velocities $v_0$ may penetrate the entire pulse, and experience the following perpendicular velocity shifts through the transit-time CRA, [1]

$$\Delta v_\perp = \frac{\sqrt{\pi}}{2} \frac{|qE_0|\Delta t}{m\gamma_0} \left[ e^{-\{\omega_0+\Omega/(\alpha\gamma_0)\}^2 \Delta t^2/2} + e^{-\{\omega_0-\Omega/(\alpha\gamma_0)\}^2 \Delta t^2/2} \right. \quad (1)$$
$$\left. + 2\cos(2\theta)e^{-\{\omega_0+\Omega/(\alpha\gamma_0)\}^2 \Delta t^2/4 - \{\omega_0-\Omega/(\alpha\gamma_0)\}^2 \Delta t^2/4} \right]^{1/2}$$

with $\alpha = 1 - v_p/v_0$, $\Delta t = l/(\beta v_p)$, $\beta = |(v_0 - v_g)/(v_0 - v_p)|$ and $\gamma_0 = (1 - v_0^2/c^2)^{-1/2}$.

Other notations are standard. We define the phase velocity as $v_p = \omega_0/k_0$. Though the theory may include dispersion effects to some extent, the present work is mainly restrained for nondispersive pulses with $v_p = v_g = 0.1c$. We will present below several numerical solutions to the electron equation of motion,[1] gradually increasing the pulse amplitude $E_0$.

## 2. Results
### 2.1 Linear Cyclotron-Resonance

We start with the linear case, which corresponds to the solution (1) above. Fig.1 depicts among others velocity shifts of electrons well after penetrating an EM wave packet with $E_n = eE_0/mc\omega_0 = 0.001$ and $l_n = l/(c/\omega_0) = 2.0$. As the magnetic field strength will be held constant at $\Omega_e = \omega_0$, where $\Omega_e > 0$ is the electron cyclotron resonance, the electrons with $v_0 \approx 0$ are resonant. Because of the finite-length pulse, the resonance velocity has a width, which is describable via Eq.(1). Unlike the usual transit-time accelerations[1] this acceleration mechanism is independent of the phase (not shown). For linear pulses perpendicular acceleration is always dominant. We find that the CRA processes depend on $\omega_t \Delta t$, where the trapping frequency equals $\sqrt{k_0 v_\perp \Omega_w}$, and $v_\perp$ is the perpendicular velocity component of the accelerated particle,[2] $\Omega_w$ is the cyclotron frequency based on the wave field, while $\Delta t$ is the transit-time; this linear regime corresponds to $\omega_t \Delta t << 2\pi$.

### 2.2 Weakly Nonlinear Cyclotron-Resonance

Fig.1 shows final velocity shifts of electrons after interactions with a linear- to weakly nonlinear wave packets with $E_n$=0.001-0.01. As the amplitude is increased, the numerical results, which show much broader acceleration region as well as multi-peaks, gradually deviate from the linear theory (1). The most outstanding nonlinear effect is the bifurcation/quantization of the linear cyclotron resonance peak, which occurs when $E_n$=0.07 in the weakly nonlinear regime such that $\omega_t \Delta t \approx 2\pi$ or greater. When $E_n$=0.01 the strongest acceleration occurs at $v_0$= -0.02$c$, while that in the linear regime occurs at $v_0 \approx 0$.

A similar trend is also observable in Fig.2, in which $E_n$ is further increased from 0.02 to 0.04. However, in this case, the increase in the amplitude tends to broaden the resonances rather than increasing the number of resonances. When $E_n$=0.04 the velocity profile is reminiscent of the phase trapping[2] that occurs in the strongly nonlinear regime. Since in all of these cases the perpendicular velocity shifts are greater than $v_p - v_0$=0.1$c$, some particles may be reflected by the pulse, and consequently the final parallel velocity of the reflected particles may reach 0.2$c$. Namely, particle reflections are a sign of strongly nonlinear regime of the CRA. In such a regime the effects of the wave potential becomes dominant. It is found also that even with the linear amplitude, as the pulse is elongated, the number of resonance peaks increases, and the band structure becomes more evident. Fig.3 shows that when the pulse with $E_n$=0.01 and $l_n$=5, which result in two resonance peaks in Fig.1, is elongated to as long as $l_n$=5 and 10 the number of resonance increases to 4 and 7, respectively. What is more, even at a low amplitude, if the pulse length is made longer such that the condition $\omega_t \Delta t \approx 2\pi$ is satisfied, the penetrating particles experience the weakly nonlinear CRA rather than the transit-time CRA, as demonstrated in Fig.4. Namely, the nonlinearity of a pulse is determined by the parameter $\omega_t \Delta t$ rather than its amplitude.

**2.3 Nonlinear pulse amplification**

Finally, Fig.5 depicts the accelerations of electrons initially with $v_o$=0.2$c$ by pulses with different amplitudes. When the pulse amplitude is small the acceleration occurs near the cyclotron frequency. As in this case, accelerated electrons lose energies, the flow of energy being reversed from the electrons to the pulse. Therefore, the components of the pulse and/or noise with $\omega \approx \Omega_e$ tend to grow. However, when the pulse amplitude is higher, the modes at $\omega \approx 1.2\Omega_e$ may grow, amplifying the pulse. Therefore, this computation shows that when a series of pulses are produced and grow to sufficiently large amplitudes via a type of cyclotron resonance instability, the frequency of the unstable modes will shift to separate ones.

**3. Conclusions**

As the pulse amplitude increases, the cyclotron resonance bifurcates, showing multi-resonance peaks. The reason for the bifurcation of resonance is that the accelerated particles experience trapping, and the resultant velocity modulation may resonate with the finite length of the pulses. It was also demonstrated that by the same resonance mechanism the combination of the pulse and particles may amplify waves at separate frequencies.

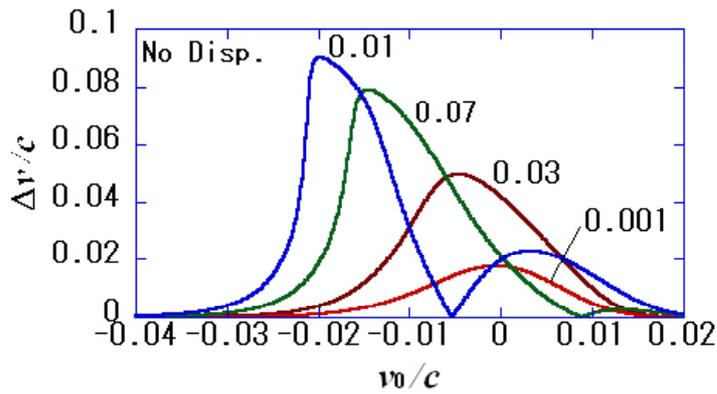

Fig.1 Perpendicular velocity shifts of electrons after penetrating an EM Gaussian pulse of various amplitudes $E_n$.

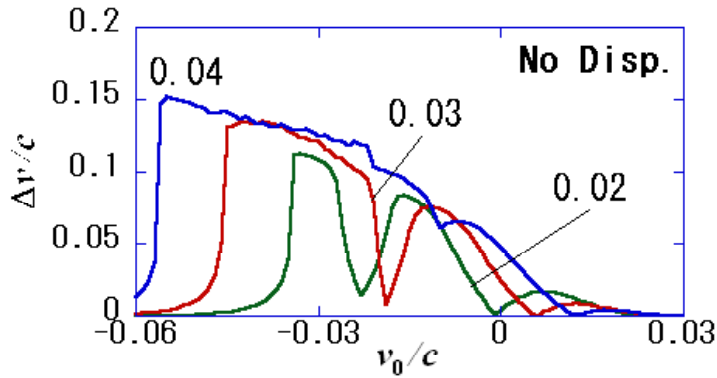

Fig.2 The same as Fig.1 for higher amplitude pulses. However, the higher amplitudes result mainly in broadening of the resonances. When $E_n=0.04$ the resonance profile looks like that of the phase-trapping.[3]

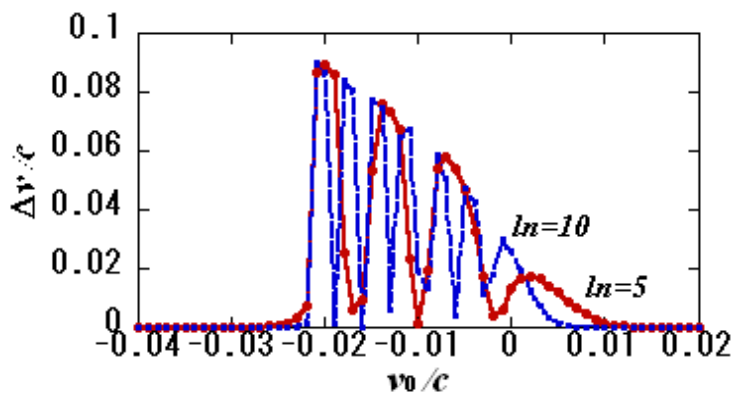

Fig.3　As the pulse ($E_n$=0.01) is elongated the number of resonances increases.

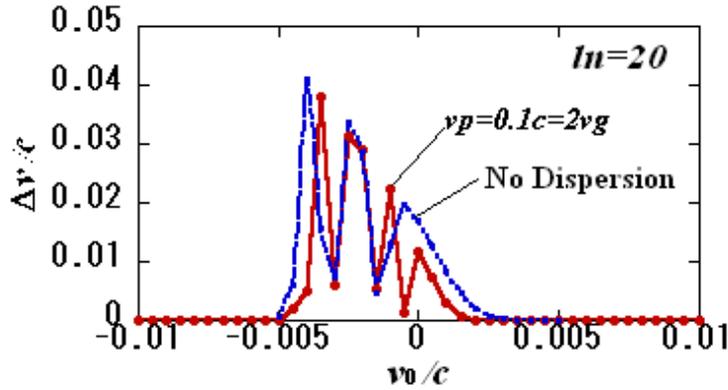

Fig.4 Even in the linear regime where the transit-time accelerations are dominant, if the pulse is elongated sufficiently, the weakly nonlinear CRA, which shows approximately quantized resonance velocities, energies. The calculations are done for nondispersive and dispersive pulses with $v_p = v_g = 0.1c$. and $v_p = 0.1c, =2v_g$, respectively.

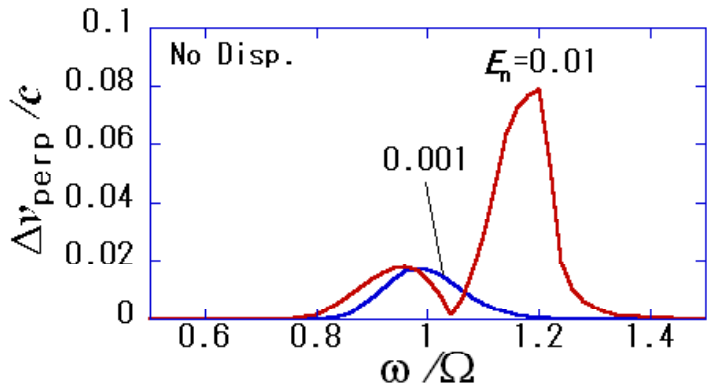

Fig.5　Frequency and $E_n$ dependencies of accelerations of electrons with $v_o$=0.2$c$ initially. When $E_n$=0.001, the acceleration due to the transit-time CRA occurs principally at $v_o$=0. Meanwhile, as the amplitude is increased to $E_n$=0.01, the active frequencies are shifted to immediately below $\Omega_e$ and $1.2\Omega_e$.